\documentclass[12pt,a4paper]{article}
\usepackage[applemac]{inputenc}
\usepackage{amssymb}
\usepackage{amsmath}
\usepackage{amsopn}
\newcommand{\LL}{\mathbb{L}}
\newcommand{\CC}{\mathbb{C}}

\newcommand{\KK}{\mathbb{K}}

\newcommand{\NN}{\mathbb{N}}

\newcommand{\RR}{\mathbb{R}}
\newcommand{\ZZ}{\mathbb{Z}}

\newcommand{\frAr}{\mathfrak{A}(\rr)}
\newcommand{\frc}{\mathfrak{C}}

\newcommand{\frb}{\mathfrak{B}}

\newcommand{\frm}{\mathfrak{m}}

\newcommand{\frsr}{\mathfrak{S}(\rr)}
\newcommand{\frU}{\mathfrak{U}}

\newcommand{\kaa}{\mathcal{A}}
\newcommand{\kbb}{\mathcal{B}}

\newcommand{\ktm}{\mathcal{T}(M)}
\newcommand{\kcc}{\mathcal{C}}

\newcommand{\cor}{\mathcal{C}\mathcal{O}\mathcal{N}(\rr)}
\newcommand{\kD}{\mathcal{D}}
\newcommand{\kE}{\mathcal{E}}
\newcommand{\kf}{\mathcal{F}}
\newcommand{\kh}{\mathcal{H}}
\newcommand{\kI}{\mathcal{I}}
\newcommand{\kj}{\mathcal{J}}

\newcommand{\kL}{\mathcal{L}}
\newcommand{\mm}{\mathcal{M}}

\newcommand{\kO}{\mathcal{O}}
\newcommand{\kP}{\mathcal{P}}
\newcommand{\kQ}{\mathcal{Q}}
\newcommand{\kR}{\mathcal{R}}
\newcommand{\kS}{\mathcal{S}}

\newcommand{\kT}{\mathcal{T}}
\newcommand{\kV}{\mathcal{V}}

\newcommand{\gb}{\beta}
\newcommand{\gd}{\delta}
\newcommand{\eps}{\varepsilon}
\newcommand{\gga}{\gamma}
\newcommand{\gG}{\Gamma}

\newcommand{\gl}{\lambda}

\newcommand{\gO}{\Omega}
\newcommand{\gf}{\varphi}

\newcommand{\gr}{\varrho}
\newcommand{\gs}{\sigma}

\newcommand{\gt}{\tau}

\newcommand{\gz}{\zeta}
\newcommand{\tm}{\subseteq}

\newcommand{\lm}{\emptyset}
\newcommand{\∞}{\infty}

\newtheorem{definition}{Definition}[section]
\newtheorem{proposition}{Proposition}[section]
\newtheorem{theorem}{Theorem}[section]
\newtheorem{lemma}{Lemma}[section]
\newtheorem{corollary}{Corollary}[section]
\newtheorem{remark}{Remark}[section]
\newtheorem{example}{Example}[section]

\newcommand{\por}{\kP_{0}(\kR)}

\newcommand{\pom}{\kP_{0}(\mm)}

\newcommand{\pr}{\kP(\kR)}
\newcommand{\ph}{\kP(\lh)}

\newcommand{\pmm}{\kP(\mm)}

\newcommand{\orr}{\kO(\kR)}
\newcommand{\omm}{\kO(\mm)}

\newcommand{\oaa}{\kO(\kaa)}

\newcommand{\qr}{\mathcal{Q}(\mathcal{R})}
\newcommand{\qm}{\kQ(\kT(M))}
\newcommand{\qpr}{\mathcal{Q}_{P}(\mathcal{R})}

\newcommand{\qh}{\mathcal{Q}(\mathcal{H})} 
\newcommand{\qa}{\mathcal{Q}(\mathcal{A})}
\newcommand{\qb}{\mathcal{Q}(\mathcal{B})}
\newcommand{\qc}{\mathcal{Q}(\mathcal{C})}

\newcommand{\ql}{\mathcal{Q}(\LL)}

\newcommand{\qal}{\mathcal{Q}_{a}(\LL)}

\newcommand{\lh}{\mathcal{L}(\mathcal{H})}
\newcommand{\llh}{\LL(\kh)}

\newcommand{\all}{\forall}
\newcommand{\ex}{\exists}
\newcommand{\rr}{\kR}

\newcommand{\hr}{\kR_{sa}}
\newcommand{\el}{E_{\gl}}
\newcommand{\jl}{\kj_{\gl}}
\newcommand{\jm}{\kj_{\mu}}
\newcommand{\emm}{E_{\mu}}

\newcommand{\eal}{E^{A}_{\gl}}

\newcommand{\ebl}{E^{B}_{\gl}}
\newcommand{\epl}{E^{P}_{\gl}}

\newcommand{\ea}{E^{A}}

\newcommand{\ef}{E^{f}}
\newcommand{\ep}{E^{P}}

\newcommand{\we}{\wedge}
\newcommand{\We}{\bigwedge}
\newcommand{\Ve}{\bigvee}

\newcommand{\tto}{\mapsto}
\newcommand{\lra}{\Longrightarrow}
\newcommand{\llra}{\Longleftrightarrow}
\newcommand{\smm}{\setminus}
\newcommand{\dl}{\kD(\LL)}
\newcommand{\dm}{\kD(\mm)}
\newcommand{\dr}{\kD(\rr)}
\newcommand{\dpr}{\kD_{P}(\rr)}
\newcommand{\dprr}{\kD_{pr}(\rr)}
\newcommand{\dal}{\kD_{a}(\LL)}
\newcommand{\dbl}{\kD_{b}(\LL)}
\newcommand{\kcb}{\gr^{\kcc}_{\kbb}}
\newcommand{\kca}{\gr^{\kcc}_{\kaa}}
\newcommand{\kba}{\gr^{\kbb}_{\kaa}}

\newcommand{\irr}{\in \RR}

\newcommand{\ikk}{\in \KK}

\newcommand{\urb}{\overset{-1}}
\newcommand{\px}{P_{\CC x}}
\newcommand{\epx}{E^{\px}}

\begin{document}

\title{\Huge{Observables}}

\author{Hans F.\ de Groote\footnote{e-mail: degroote@math.uni-frankfurt.de}
 \\ FB Mathematik\\ J.W.Goethe-Universität\\ Frankfurt a.\ M.}

\date{June 10, 2005\footnote{Talk, given at the colloquium ``New 
mathematical structures in the foundations of quantum theory''}}
\maketitle

\begin{quote}
    \emph{Ein etwas vorschnippischer Philosoph, ich glaube Hamlet, Prinz von
    Dänemark, hat gesagt, es gebe eine Menge Dinge im Himmel und auf
    der Erde, wovon nichts in unseren Compendiis steht. Hat der
    einfältige Mensch, der bekanntlich nicht recht bei Trost war,
    damit auf unsere Compendia der Physik gestichelt, so kann man ihm 
    getrost antworten: Gut, aber dafür stehn auch wieder eine Menge
    von Dingen in unseren Compendiis, wovon weder im Himmel noch auf
    der Erde etwas vorkömmt. \\
    (Georg Christoph Lichtenberg, Aphorismen 1796 - 1799)}            
\end{quote}
\vspace{2cm}

Is there any justification to speak nearly an hour about such a basic 
and apparently simple notion like ``observable''? Are there some new -
and interesting - insights, at least mathematically, or even conceptually?
\\

I propose that this is the case. \\

Following Araki \cite{ar}, an observable is an equivalence class of
measuring instruments, two measuring instruments being equivalent if
in any ``state'' of the ``physical system'' they lead upon a ``large
number of measurements'' to the same distribution of (relative
frequencies of) results. From this concept one can derive (a little
bit in the sense of Ed Harris) that
\begin{itemize}
    \item [(i)] in (von Neumann's axiomatic approach to) quantum physics,
    an observable is represented by a bounded selfadjoint operator $A$
    acting on a Hilbert space $\kh$, and  

    \item [(ii)] in classical mechanics, an observable is represented by a
    real valued (smooth or continuous or measurable) function on an
    appropriate phase space.
\end{itemize}
Here a natural question arises: is the structural difference
between classical and quantum observables fundamental, or is there
some background structure showing that classical and quantum observables
are on the same footing? Indeed, such a background structure exists,
and I shall describe in this talk some of its features. 

\section{Motivation: Presheaves on a Lattice}
\label{mo}

A continuous classical observable is a continuous function $f : M \to 
\RR$ on a (locally compact) Hausdorff space $M$. Equivalently, $f$ can
be considered as a global section of the presheaf $\kcc_{M}$ of all 
real valued continuous functions that are defined on some nonempty
open subsets of $M$. This situation leads to a natural generalization.
The set $\ktm$ of all open subsets of $M$ can be seen as a
\emph{complete lattice}, the lattice operations being defined by
\[
    \Ve_{k \ikk}U_{k} := \bigcup_{k \ikk}U_{k}, \qquad \We_{k \ikk}U_{k}
    := int (\bigcap_{k \ikk}U_{k}). 
\] 
(In the English language the word ``lattice'' has two different
meanings. Either it is a subgroup of the additive group $\ZZ^d$ for
some $d \in \NN$ (this is called ``Gitter'' in German) or it means a
partially ordered set with certain additional properties. This is
called ``Verband'' in German. We always use ``lattice'' in this second
meaning.) \\

\begin{definition}\label{mo1}
A \emph{lattice} is a partially ordered set $(\LL, \leq)$  such 
that any two elements $a,b \in\LL$ possess a \emph{maximum} $a\vee b 
\in\LL$ and a \emph{minimum} $a\wedge b \in\LL$.\\
Let $\frm$ be an infinite cardinal number.\\
The lattice $\LL$ is called $\mathfrak{m}$-complete, if every family 
$(a_{i})_{i\in I}$ has a supremum $\bigvee_{i \in I}a_{i}$ and an 
infimum $\bigwedge_{i\in I}a_{i}$ in $\LL$, provided that $\# I\leq 
\mathfrak{m}$ holds.
A lattice $\LL$ is simply called complete, if every family 
$(a_{i})_{i \in I}$ in $\LL$ (without any restriction of the 
cardinality of $I$) has a supremum and an infimum in $\LL$.\\
$\LL$ is said to be boundedly complete if every bounded family in 
$\LL$ has a supremum and an infimum. \\
If a lattice has a \emph{zero element} $0$ ( i.e. $\forall a \in \LL: 
0 \leq a$) and a \emph{unit element} $1$ (i.e. $\forall a\in \LL : a\leq 
1$) then completeness and bounded completeness are the same.\\
A lattice $\LL$ is called \emph{distributive} if the two distributive 
laws
\begin{eqnarray*}
       a \wedge (b \vee c) & = & (a \wedge b) \vee (a \wedge c)  \\
       a \vee (b \wedge c) & = & (a \vee b) \wedge (a \vee c)
\end{eqnarray*}
hold for all elements $a, b, c \in \LL$. 
\end{definition}

In fact it is an easy exercise to show that if one of these
distributive laws is satisfied for all $a, b, c \in \LL$ so is the
other. \\

%


Traditionally, the notions of a {\bf presheaf} and a {\bf complete 
presheaf} (complete presheaves are usually called ``sheaves'') are 
defined for the lattice $\kT(M)$ of a topological space $M$. The very 
definition of presheaves and sheaves, however, can be formulated also 
for an arbitrary lattice:
\begin{definition}\label{mo2}
A {\bf presheaf} of sets ($R$-modules) on a lattice $\LL$ assigns to 
every element $a \in \LL$ a set ($R$-module) $\kS(a)$ and to every 
pair $(a, b) \in \LL \times \LL$ with $a \leq b$ a mapping 
($R$-module homomorphism)
\begin{displaymath}
	\rho_{a}^{b} : \kS(b) \to \kS(a)
\end{displaymath}
such that the following two properties hold:
\begin{enumerate}
	\item  [(1)] $\rho_{a}^{a} = id_{\kS(a)}$ for all $a \in \LL$,

	\item  [(2)] $\rho_{a}^{b} \circ \rho_{b}^{c} = \rho_{a}^{c}$ for 
	all $a, b, c \in \LL$ such that $a \leq b \leq c$.
\end{enumerate}
The presheaf $(\kS(a), \rho_{a}^{b})_{a \leq b}$ is called a {\bf 
complete presheaf} (or a {\bf sheaf} for short) if it has the 
additional property
\begin{enumerate}
	\item  [(3)] If $a = \bigvee_{i \in I}a_{i}$ in $\LL$ and if $f_{i} 
	\in \kS(a_{i}) \ \ (i \in I)$ are given such that
	\begin{displaymath}
		\forall \ i, j \in I : \ (a_{i} \wedge a_{j} \ne 0 \quad 
	\Longrightarrow \quad \rho_{a_{i}\wedge a_{j}}^{a_{i}}(f_{i}) = 
	\rho_{a_{i} \wedge a_{j}}^{a_{j}}(f_{j}),
	\end{displaymath}
	then there is exactly one $f \in \kS(a)$ such that
	\begin{displaymath}
		\forall \ i \in I : \ \rho_{a_{i}}^{a}(f) = f_{i}.
	\end{displaymath}
\end{enumerate}
The mappings $\rho_{a}^{b} : \kS(b) \to \kS(a)$ are called {\bf 
restriction maps}.
\end{definition}

Concerning quantum theory, the fundamental lattice is the lattice
$\llh$ of all \emph{closed} linear subspaces of a Hilbert space $\kh$.
This lattice is defined by
\begin{enumerate}
    \item  [(i)] $U ≤ V \quad :\llra \quad U \tm V$,

    \item  [(ii)] $\We_{k \ikk}U_{k} := \bigcap_{k \ikk}U_{k}$,

    \item  [(iii)] $\Ve_{k \ikk}U_{k} := \overline{\sum_{k
    \ikk}U_{k}}$.
\end{enumerate}
There are interesting presheaves on $\llh$ but, unfortunately, any
complete presheaf on $\llh$ is also completely trivial:

\begin{proposition}\label{mo3}
Let $(\kS(U), \rho_{U}^{V})_{U \tm V}$ be a complete presheaf of
nonempty sets on the quantum lattice $\LL(\kh)$. Then
\begin{displaymath}
	\#\kS(U) = 1
\end{displaymath}
for all $U \in \LL(\kh)$.
\end{proposition}
In order to present a fundamental example of a presheaf on $\llh$
(which turns out to be rather universal), we make the following general

\begin{definition}\label{mo4}
    Let $\LL$ be a complete lattice. A mapping
    \[
        \begin{array}{cccc}
            E : & \RR & \to & \LL  \\
             & \gl & \tto & \el
        \end{array}
    \]
    is called a {\bf spectral family} in $\LL$, if it has the
    following properties:
    \begin{enumerate}
        \item  [(i)] $\el ≤ \emm$ for $\gl ≤ \mu$,
    
        \item  [(ii)] $\el = \We_{\mu > \gl}\emm$ for all $\gl \irr$,
    
        \item  [(iii)] $\We_{\gl \irr}\el = 0, \qquad \Ve_{\gl
	\irr}\el = 1$.
    \end{enumerate}
    A spectral family $E$ in $\LL$ is called bounded if there are $a, 
    b \irr$ such that $\el = 0$ for all $\gl < a$ and $\el = 1$ for
    all $\gl ≥ b$.  
\end{definition}
If $\LL$ is a complete lattice and $a \in \LL$, then 
\[
    \LL_{a} := \{ b \in \LL \ | \ b ≤ a \}
\]
is a complete sublattice with maximal element $a$. Let $E$ be a
bounded spectral family in $\LL$. Then
\[
    E^{a} : \gl \tto \el \we a
\] 
is a spectral family in $\LL_{a}$. $E^{a}$ is called the
\emph{restriction of $E$ to $a$}. Let $\kE(a)$ be the set of all
spectral families in $\LL_{a}$. Then for $a, b \in \LL, \ a ≤ b,$ we
obtain a mapping
\[
    \begin{array}{cccc}
        \gr_{ab} : & \kE(b) & \to & \kE(a)  \\
         & E & \tto & E^{a},
    \end{array}
\] 
and it is easy to see that the sets $\kE(a)$ together with the
restriction maps $\gr_{ab}$ form a presheaf on $\LL$. We denote this
presheaf by $\kE_{\LL}$ and call it the \emph{spectral presheaf on
$\LL$}. Let us give a ``classical'' example:
\begin{example}\label{mo5}
    Let $f : M \to \RR$ be a continuous function on a Hausdorff space 
    $M$. Then 
    \[
        \all \ \gl \irr : \ \el := int (\urb{f}(]-\∞, \gl ]))
    \]
    defines a spectral family $E : \RR \to \ktm$. (The natural
    guess for defining a spectral family corresponding to $f$ would be
    \[
        \gl \tto \urb{f}(]-\∞, \gl [).
    \]
    In general, this is only a \emph{pre-spectral family}: it
    satisfies all properties of a spectral family, except continuity
    from the right. This is cured by \emph{spectralization}, i.e. by
    the switch to
    \[
        \gl \tto \We_{\mu > \gl}\urb{f}(]-\∞, \mu [).
    \]
    But
    \[
    \We_{\mu > \gl}\urb{f}(]-\∞, \mu [) = int (\bigcap_{\mu > \gl}
    \urb{f}(]-\∞, \mu [)) = int (\urb{f}(]-\∞, \gl ])),
    \]
    which shows that our original definition is the natural one.)\\
    One can show that
    \[
        \all \ x \in M : \ f(x) = \inf \{ \gl \ | \ x \in \el \},
    \]
    so one can recover the function $f$ from its spectral family $E$. 
    Let $U \in \ktm, \ U \ne \emptyset$. Then 
    \[
        \el \cap U = int \{ x \in U \ | \ f(x) ≤ \gl \} =
	int(\urb{f_{|_{U}}}(]-\∞, \gl])),  
    \]
    hence $E^{U}$ is the spectral family of the usual restriction
    $f_{|_{U}} : U \to \RR$ of $f$ to $U$. 
\end{example}
Another example comes from the theory of operator algebras:

\begin{example}\label{mo6}
    Let $\rr$ be a von Neumann algebra, acting on a Hilbert space
    $\kh$, let $\hr$ be the set of selfadjoint operators in $\rr$ and let
    $\pr$ be the lattice of projections in $\rr$. The spectral family 
    $E$ of $A \in \hr$ is a spectral family in $\pr$ and therefore the
    restriction
    \[
        E^{P} : \gl \tto \el \we P
    \]
    of $E$ to $P \in \pr$ is a bounded spectral family in the
    \emph{ideal}
    \[
        I_{P} :=  \{ Q \in \pr \ | \ Q ≤ P \} \tm \pr.
    \]
    It is easy to see that the operator $A^{P}$ corresponding to
    $E^{P}$ belongs to $P\rr P$, a von Neumann algebra operating on
    the Hilbert space $P\kh$. In particular, $A^{P}$ can be considered
    as an element of $\kL(P\kh)$. Note that, although $A^{P} \in P\rr 
    P$, $A^{P}$ is, in general, different from $PAP$. This follows
    from the fact that $A^{P}$ is a projection if $A$ is, but that,
    for a projection $A$, $PAP$ is a projection if and only if $A$
    commutes with $P$. \\
    For $P \in \pr$, let $\kE(P)$ be the set of bounded spectral
    families in $I_{P}$, and for $P, Q \in \pr$ such that $P ≤ Q$, we 
    define a restriction map
    \[
        \begin{array}{cccc}
            \gr_{PQ} : & \kE(Q) & \to & \kE(P)  \\
             & E & \tto & E^{P}.
        \end{array}
    \]       
    It is obvious that the sets $\kE(P)$ together with the restriction
    maps $\gr_{PQ}$ form a presheaf $\kE_{\rr}$ on the projection
    lattice $\pr$. \\
    In general, one cannot drop the assumption that the spectral
    families are bounded, for it may happen that, if $E$ is not
    bounded from above, $\Ve_{\gl \irr}\epl \ne P$ holds. More
    precisely, one can show (\cite{dg}) that $\ep$ is a spectral family
    for all $P \in \pr$ and any spectral family $E$ in $\pr$ if and
    only if $\rr$ is a \emph{finite} von Neumann algebra. 
\end{example}
It is well known that one can associate to each preasheaf $\kS_{M}$ on a
topological space $M$ a sheaf on $M$ in the following way: \\
If $\kS$ is a presheaf on a topological space $M$, 
i.e. on the lattice $\kT(M)$, then the corresponding \emph{etale
space} $\kE(\kS)$ of $\kS$ is the disjoint union of the \emph{stalks}
of $\kS$ at points in $M$:
\[  \kE(\kS) = \coprod_{x \in M}\kS_{x} \]
where
\[  \kS_{x} = \lim_{\overset{\longrightarrow}{U \in \frU}}\kS(U), \]
the \emph{inductive limit} of the family $(\kS(U))_{U \in \frU(x)}$,
(here $\frU(x)$ denotes the set of all open neighbourhoods of $x$)
is the stalk in $x \in M$. The stalk $\kS_{x}$ consists of the
\emph{germs} in $x$ of elements $f \in \kS(U), \ U \in \frU(x)$. Germs
are defined quite analogously to the case of ordinary functions. Let
$\pi : \kE(\kS) \to M$ be the mapping that sends a germ in $x$ to its 
basepoint $x$. $\kE(\kS)$ can be given a topology for which $\pi$
is a local homeomorphism. It is easy to see that the local sections of
$\pi$ form a complete presheaf on $M$. If $\kS$ was already complete, 
then this presheaf of local sections of $\pi$ is isomorphic to $\kS$.\\

A first attempt to generalize this construction to the situation of a 
presheaf on a general lattice $\LL$ is to define a suitable notion of
``point in a lattice''. This can be done in a quite natural manner,
and it turns out that, for \emph{regular} topological spaces $M$,
the points in $\ktm$ are of the form $\frU(x)$, hence correspond to
the elements of $M$. But it also turns out that some important
lattices, like $\llh$, do not have points at all (\cite{dg})! \\  

For the definition of an inductive limit, however, we do not need a 
point, like $\frU(x)$, but only a partially ordered set $I$ with the property
\[  \forall \ \alpha, \beta \in I\ \  \exists \gamma \in I: \gamma \leq 
\alpha \ \  \mbox{and}  \ \ \gamma \leq \beta. \]
In other words: a \emph{filter base} $B$ in a lattice $\LL$ is 
sufficient. It is obvious how to define a filter base in an 
arbitrary lattice $\LL$:

\begin{definition}\label{mo7} A filter base $B$ in a lattice $\LL$ is a 
non-empty subset $B \subseteq \LL$ such that
\begin{enumerate}
	\item  [(1)] $0 \notin B$,

	\item  [(2)] $\forall \ a,b \in B \ \exists \ c \in B : \ c \leq a 
	\wedge b$.
\end{enumerate}
\end{definition}

The set of all filter bases in a lattice $\LL$ is of course a vast object. 
So it is reasonable to consider \emph{maximal} filter bases in 
$\LL$. (By Zorn's lemma, every filter base is contained in a maximal 
filter base in $\LL$.) This leads to the following

\begin{definition}\label{mo8} A nonempty subset $\frb$ of a lattice $\LL$ is 
called a {\bf quasipoint} in $\LL$ if and only if
\begin{enumerate}
	\item  [(1)] $0 \notin \frb$,

	\item  [(2)] $\forall \ a,b \in \frb \ \exists \ c \in \frb : \ c 
	\leq a \wedge b$,

	\item  [(3)] $\frb$ is a maximal subset having the properties $(1)$ 
	and $(2)$.
	\end{enumerate}
	We denote the set of quasipoints in $\LL$ by $\ql$.
\end{definition}
It is easy to see that a quasipoint in $\LL$ is nothing else but a
\emph{maximal dual ideal}.

In 1936 M.H.Stone (\cite{stone}) showed that the set $\kQ(\kbb)$ of 
quasipoints in a Boolean algebra $\kbb$ can be given a topology such 
that $\kQ(\kbb)$ is a \emph{compact zero dimensional} Hausdorff space 
and that the Boolean algebra $\kbb$ is isomorphic to the Boolean 
algebra of all \emph{closed open} subsets of $\kQ(\kbb)$. A basis for 
this topology is simply given by the sets
\[ \kQ_{U}(\kbb) := \{ \frb \in \kQ(\kbb) \mid U \in \frb \} \]
where $U$ is an arbitrary element of $\kbb$. \\

Of course we can generalize this construction to an arbitrary 
lattice $\LL$:\\
For $a \in \LL$ let
\[ \kQ_{a}(\LL) := \{ \frb \in \kQ(\LL) \mid a \in \frb \}. \]
It is quite obvious from the definition of a quasipoint that
\[ \kQ_{a \wedge b}(\LL) = \kQ_{a}(\LL) \cap \kQ_{b}(\LL), \]
\[\kQ_{0}(\LL) = \emptyset \quad \text{and} \quad \kQ_{I}(\LL)
    = \kQ(\LL)  \]
hold. Hence $\{ \kQ_{a}(\LL) \mid a \in \LL \}$ is a basis for a 
topology on $\kQ(\LL)$. It is easy to see, using the 
maximality of quasipoints, that in this topology the sets 
$\kQ_{a}(\LL)$ are open and closed. Moreover, this topology is
Hausdorff, zero-dimensional, and therefore also completely regular. \\

\begin{definition}\label{mo9}
    $\kQ(\LL)$, together with the topology defined by the basis $\{
    \qal \ | \ a \in \LL \}$, is called the {\bf Stone spectrum
    of the lattice $\LL$.}
\end{definition}

Then we can mimic the construction of the etale space of a presheaf on
a topological space $M$ to obtain from a presheaf $\kS$ on a lattice
$\LL$ an etale space $\kE(\kS)$ \emph{over the Stone spectrum $\ql$}
and a local homeomorphism $\pi_{\kS} : \kE(\kS) \to \ql$.
From the etale space $\kE(\kS)$ over $\kQ(\LL)$ we obtain 
a complete presheaf $\kS^{\kQ}$ on the topological space $\kQ(\LL)$ by
  \[ \kS^{\kQ}(\kV) := \gG(\kV,\ \kE(\kS)) \]
where $\kV \subseteq \kQ(\LL)$ is an open set and $\gG(\kV,\ 
\kE(\kS))$ is the set of {\bf sections of $\pi_{\kS}$ 
over $\kV$}, i.e. of all (necessarily continuous) mappings $s_{\kV} : \kV \to 
\kE(\kS)$ such that $\pi_{\kS} \circ s_{\kV} = id_{\kV}$. If $\kS$ 
is a presheaf of modules, then $\gG(\kV, \kE(\kS))$ is a module,
too.\\

\begin{definition}\label{mo10}
   The complete presheaf $\kS^{\kQ}$ on the Stone spectrum $\kQ(\LL)$ 
   is called the {\bf sheaf associated to the presheaf $\kS$ on $\LL$}.
\end{definition}

Of course, Stone had quite another motivation for introducing the
space $\qb$ of a Boolean algebra $\kbb$, namely to represent $\kbb$ as
a Boolean algebra of sets. The remarkable fact is that we arrive at a
generalization of Stone's concept from a completely different point of
view.\\

Let us return to example \ref{mo6} and consider the special case $\rr 
= \lh$. If $x \in \kh$ is a unit vector, $\px$ the orthogonal projection
onto the line $\CC x$, $A \in \lh$ is selfadjoint and if $E$ is its spectral
family, then the restriction $\epx$ of $E$ to $\px$ corresponds to a
linear operator
\[
    A^{\px} : \CC x \to \CC x
\]     
which is a scalar multiple $cI_{\CC x}$ of the identity $I_{\CC x}$.
Because $\kP(\CC x) = \{0, \px\}$, the spectral family $\epx$ has the 
form
\begin{displaymath}
    \epx_{\gl} =
    \begin{cases}
    0     &\text{for $\gl < c$}\\
    \px  &\text{for $\gl \geq c$}.
    \end{cases}
\end{displaymath}
It is obvious from the definition of restriction that
\begin{equation}
    c = \inf \{ \gl \ | \ \px ≤ \el \}.
    \label{eq:mo1}
\end{equation}
This is the right place to report some results on Stone spectra of the
projection lattice of a von Neumann algebra.\\
If $\rr$ is an abelian von Neumann algebra, then $\rr$ is
$\ast$-isomorphic to the von Neumann algebra $C(\gO)$, the algebra of 
all complex-valued continuous functions on the \emph{Gelfand spectrum}
$\gO$ of $\rr$. $\gO$ is the set of all multiplicative positive linear 
functionals on $\rr$. It is a compact space with respect to the
weak*-topology. 

\begin{theorem}\label{mo11}
    Let $\rr$ be an abelian von Neumann algebra. Then the Gelfand spectrum
    $\gO$ is homeomorphic to the Stone  spectrum $\qr$ of $\rr$.
\end{theorem}
The proof is based on the simple observation that, for every $\gt \in 
\gO$, 
\[
    \gb(\gt) := \{ P \in \pr \ | \ \gt(P) = 1 \}
\]
is a quasipoint in $\pr$. The mapping $\gt \tto \gb(\gt)$ turns out to
be the assured homeomorphism. \\

\begin{proposition}\label{mo12}
Let $\kh$ be a Hilbert space and let $\frb$ be a quasipoint in $\ph$.
$\frb$ contains an element of finite rank if and only if there is a
(necessarily unique) line $\CC x$ in $\kh$ such that
\[ \frb = \{ P \in \ph \ \mid \px ≤ P \} . \]
If $\kh$ has infinite dimension, then $\frb$ does not contain an element
of finite rank if and only if $P \in \frb$ for all $P \in \ph$ of finite corank.
\end{proposition}

Quasipoints of the form $\{ P \in \ph \ \mid \px ≤ P \}$ in $\ph$ are 
called \emph{atomic}. These are precisely the \emph{isolated points}
of the topological space $\qh$. They form a dense subset of $\qh$.\\

Let $\rr$ be an arbitrary von Neumann algebra with center $\kcc$. It
can be shown that, for every $\frb \in \qr$, the intersection $\frb
\cap \kcc$ is a quasipoint in $\kcc$, and that the mapping
\[
    \begin{array}{cccc}
        \gz : & \qr & \to & \qc  \\
         & \frb & \tto & \frb \cap \kcc 
    \end{array}
\]  
is surjective, continuous and open.\\
The unitary group $\frU_{\rr}$ of $\rr$ acts in a natural way on
the Stone spectrum $\qr$:
\[
    T.\frb := \{ T^{\ast}PT  \mid P \in \frb \} \in \qr 
\] 
for all $T \in \frU_{\rr}$ and all $\frb \in \qr$.

\begin{theorem}(\cite{doe})\label{mo13}
    Let $\rr$ be a finite von Neumann algebra  of type $I$ and let
    $\kcc$ be the center of $\rr$. Then the orbits of the action of
    the unitary group $\frU_{\rr}$ of $\rr$ on $\qr$ are the fibres
    of $\gz$.
\end{theorem}
The characterization of the Stone spectra for other types of von
Neumann algebras seems to be a really hard problem. \\

Note that equation (\ref{eq:mo1}) can be rewritten as
\[
    c = \inf \{ \gl \mid \el \in \frb_{\px} \}, 
\]
where $\frb_{\px}$ is the atomic quasipoint determined by $\px$. This 
gives a function on the set of atomic quasipoints of $\ph$ which has a
natural extension to the whole Stone spectrum $\qh$:
\[
    \all \ \frb \in \qh : \ f_{A}(\frb) := \inf \{ \gl \mid \el \in \frb \}.
\]   
Of course this can be generalized to arbitrary von Neumann algebras
and even to arbitrary complete lattices:

\begin{definition}\label{mo14}
    Let $E : \RR \to \LL$ be a bounded spectral family in a complete
    lattice $\LL$. Then the function
    \[
        \begin{array}{cccc}
            f_{E} : & \ql & \to & \RR  \\
             & \frb & \tto & \inf \{ \gl \mid \el \in \frb \}
        \end{array}
    \] 
    is called the {\bf observable function corresponding to $E$.} 
\end{definition}

\section{Observable Functions}
\label{ob}

In this section we discuss quantum and classical observables from a
new perspective. 

\subsection{Quantum Observables}

In what follows, $\rr$ is a von Neumann algebra acting on a Hilbert
space $\kh$. $\hr$ denotes the set of all selfadjoint elements of
$\rr$. If $A \in \hr$ with spectral family $E$, then the observable
function $f_{A} : \qr \to \RR$ is defined by
\[
    f_{A}(\frb) := \inf \{ \gl \mid \el \in \frb \}.
\]
$sp(A)$ denotes the spectrum of $A$.

\subsubsection{Basic Properties}

The first basic property of observable functions is

\begin{theorem}\label{ob1}
    Let $A \in \hr$ and let $f_{A} : \qr \to \RR$ be the observable
    function corresponding to $A$. Then 
    \[
	im f_{A} = sp(A).
    \] 
\end{theorem}
The proof rests on the fact that the spectrum $sp(A)$ of $A$
consists of all $\gl \in \RR$ such that the spectral family $E$ of $A$
is non-constant on every neighbourhood of $\gl$.\\

\begin{example}\label{ob2}
    The observable function of a projection $P \in \pr$ is given by
    \[
	f_{P} = 1 - \chi_{\kQ_{I - P}(\rr)},
    \]
    where $\chi_{\kQ_{I - P}(\rr)}$ denotes the characteristic
    function of the open closed set $\kQ_{I - P}(\rr)$. Hence $f_{P}$ 
    is a continuous function. \\
    If $\rr$ is abelian then
    \[
        f_{P} = \chi_{\qpr}.
    \]
\end{example}
This example is not accidental:

\begin{theorem}\label{ob3}
    Let $A \in \hr$. Then the observable function $f_{A} : \qr \to
    \RR$ is continuous.
\end{theorem}
The proof requires some work: we know from the spectral theorem that
$A$ can be approximated in norm by finite linear combinations of
pairwise orthogonal projections. Explicit calculation of the
observable function of such a linear combination shows that it is
continuous. One proves eventually that the observable functions of the
approximating operators converge uniformly to $f_{A}$. 

\begin{definition}\label{ob4}
    Let $\rr$ be a von Neumann algebra. Then we denote by $\orr$
    the set of observable functions $\qr \to \RR$.
\end{definition}

By the foregoing result $\orr$ is a subset of $C_{b}(\qr, \RR)$, the
algebra of all bounded continuous functions $\qr \to \RR$. $\orr$
separates the points of $\qr$ because the observable function of a
projection $P$ is $f_{P} = 1 - \chi_{\kQ_{I - P}(\rr)}$. Moreover it
contains the constant functions. In general, however, it is not an 
algebra and not even a vector space (with respect to the pointwise
defined algebraic operations).

\begin{theorem}\label{ob5}
    Let $\rr$ be a von Neumann algebra and let $\orr$ be the set of
    observable functions on $\qr$. Then 
    \[
	\orr = C_{b}(\qr, \RR)
    \]
    if and only if $\rr$ is {\bf abelian}.
\end{theorem}

The fact that $\orr = C(\qr, \RR)$ for abelian $\rr$ (note that $\qr$ 
is compact in this case) has a deeper reason:

\begin{theorem}\label{ob6}
    Let $\kaa$ be an abelian von Neumann algebra. Then the mapping
    $A \tto f_{A}$ from $\kaa$ onto $C(\qa, \RR)$ is, up to the
    homeomorphism of theorem \ref{mo11}, the restriction 
    of the Gelfand transformation to $\kaa_{sa}$.
\end{theorem}
\emph{This result shows that the concept of ``observable function'' is
a noncommutative extension of the concept of ``Gelfand transform''.}\\

Here a natural question arises: Can we characterize abstractly those
elements of $C_{b}(\qr, \RR)$ that are observable functions?
An ``abstract characterization'' of an observable function $f_{A}$
should be a set of properties of $f_{A}$ in which the operator $A$
does not occur explicitely.

\subsubsection{Abstract Characterization of Observable Functions}

Let $A$ be a selfadjoint element of a von Neumann algebra $\rr$. The
definition of the observable function $f_{A} : \qr \to \RR$,
\[
      f_{A}(\frb) := \inf \{ \gl \mid \el \in \frb \},
\]
can be extended without any change to the space of \emph{all} dual
ideals in $\pr$.

\begin{definition}\label{ob7}
    Let $\LL$ be a complete lattice (with minimal element $0$ and
    maximal element $1$). A nonempty subset $\kj \tm \LL$ is called a 
    {\bf dual ideal} if it has the following properties:
    \begin{enumerate}
	\item  [(i)] $0 \notin \kj$,
    
	\item  [(ii)] $a, b \in \kj \ \lra \ a \we b \in \kj$,
    
	\item  [(iii)] if $a \in \kj$ and $a ≤ b$ then $b \in \kj$.
    \end{enumerate}
    If $a \in \LL \smm \{0\}$ then the dual ideal
    \[
	H_{a} := \{ b \in \LL | \ b ≥ a \}
    \]
    is called the {\bf principal dual ideal generated by $a$}.
    We denote by $\dl$ the set of dual ideals of $\LL$. For $a \in \LL$ let
    \[
	\dal := \{ \kj \in \dl | \ a \in \kj \}.
    \]
\end{definition}
As mentioned earlier a maximal dual ideal is nothing but a quasipoint 
of $\LL$. We collect some obvious properties of the sets $\dal$ in
the following
\begin{remark}\label{ob8}
    For all $a, b \in \LL$ the following properties hold:
    \begin{enumerate}
	\item  [(i)] $a ≤ b \ \lra \ \dal \tm \dbl$, 
    
	\item  [(ii)] $ \kD_{a \we b}(\LL) = \dal \cap \dbl$,
    
	\item  [(iii)] $\dal \cup \dbl \tm \kD_{a \vee b}(\LL)$,
    
	\item  [(iv)] $\kD_{0}(\LL) = \lm, \ \kD_{1}(\LL) = \dl$.
    \end{enumerate}
    These properties show in particular that $\{ \dal | \ a \in \LL
    \}$ is a basis of a topology on $\dl$. The Stone spectrum $\ql$ is
    dense in $\dl$ with respect to this topology.
\end{remark}
Note that $\dl$ is in general {\bf not a Hausdorff space}: let $b, c
\in \LL, b < c$. Then $H_{c} \in \dal \ \llra \ c ≤ a$, hence $H_{b} 
\in \dal$ and therefore $H_{b}$ and $H_{c}$ cannot be separated. We
return to the case that $\LL$ is the projection lattice $\pr$ of a von
Neumann algebra $\rr$, although most of our considerations also hold
for an arbitrary orthocomplemented complete lattice.\\

We shall need the following simple

\begin{lemma}\label{ob9}
    $\all \ P \in \pr : \ H_{P} = \bigcap_{P \in \frb}\frb \ \  (=: \bigcap
    \qpr)$.
\end{lemma}

Let $A \in \hr$ with corresponding spectral family $\eal$ and observable
function $f_{A}$. We extend $f_{A}$ to a function $\dr \to \RR$ on the
space $\dr$ of dual ideals of $\pr$ (and we denote this extension again
by $f_{A}$) in a natural manner:
\[
    \all \ \kj \in \dr : \ f_{A}(\kj) := \inf \{ \gl | \ \eal \in \kj
    \}.
\]
There are two fundamental properties of the function $f_{A} : \dr \to 
\RR$. The first one is expressed in the following

\begin{proposition}\label{ob10}
    Let $(\kj_{j})_{j \in J}$ be a family in $\dr$. Then
    \[
	f_{A}(\bigcap_{j \in J}\kj_{j}) = \sup_{j \in
	J}f_{A}(\kj_{j}).
    \]
\end{proposition}
The other is

\begin{proposition}\label{ob11}
    $f_{A} : \dr \to \RR$ is upper semicontinuous, i.e.
\[
    \all \ \kj_{0} \in \dr \ \all \ \eps > 0 \ \ex \ P \in \kj_{0} \
    \all \ \kj \in \dpr : \ f_{A}(\kj) < f_{A}(\kj_{0}) + \eps. 
\]
\end{proposition}
Note that, even if $A$ is a projection, $f_{A} : \dr \to \RR$ need not
be continuous!\\
Proposition \ref{ob10} implies that $f_{A}$ is \emph{decreasing},
where $\dr$ is partially ordered by inclusion. Upper semicontinuity
for decreasing functions $f : \dr \to \RR$ can be reformulated:

\begin{proposition}\label{ob12}
    For any function $f : \dr \to \RR$ the following two properties are
    equivalent:
    \begin{enumerate}
	\item  [(i)] $f$ is upper semicontinuous and decreasing.  
    
	\item  [(ii)] $\all \ \kj \in \dr  : \ f(\kj) = \inf \{
	f(H_{P}) | \ P \in \kj \}$.
    \end{enumerate}
\end{proposition}

A direct consequence of this, proposition \ref{ob10} and of lemma
\ref{ob9} is
\[
    \all \ \kj \in \dr : \ f_{A} (\kj) = \inf_{P \in \kj}\sup_{\frb
    \in \qpr}f_{A}(\frb)
\]
and, in particular, $f_{A}(\dr) = sp(A)$ for all $A \in \hr$.

\begin{definition}\label{ob13}
    A function $f : \dr \to \RR$ is called an {\bf abstract observable
    function} if it is \emph{upper semicontinuous} and satisfies the
    \emph{intersection condition}
    \[
	f(\bigcap_{j \in J}\kj_{j}) = \sup_{j \in J}f(\kj_{j}) 
    \]
    for all families $(\kj_{j})_{j \in J}$ in $\dr$.
\end{definition}

The intersection condition implies that an abstract observable
function is decreasing. Hence by proposition \ref{ob12} the definition
of abstract observable functions can be reformulated as follows:

\begin{remark}\label{ob14}
    $f : \dr \to \RR$ is an observable function if and only if the
    following two properties hold for $f$:
    \begin{enumerate}
           \item  [(i)]  $\all \ \kj \in \dr  : \ f(\kj) = \inf \{ f(H_{P}) | \ P \in \kj
	\}$,
    
           \item  [(ii)] $f(\bigcap_{j \in J}\kj_{j}) = \sup_{j \in J}f(\kj_{j})$
	for all families $(\kj_{j})_{j \in J}$ in $\dr$.
    \end{enumerate}
\end{remark}

A direct consequence of the intersection condition is the following

\begin{remark}\label{ob15}
    Let $\gl \in im f$. Then the inverse image $\overset{-1}{f}(\gl) \tm
    \dr$ has a minimal element $\kj_{\gl}$ which is simply given by
    \[
	\kj_{\gl} = \bigcap \{ \kj \in \dr | \ f(\kj) = \gl \}.
    \]   
\end{remark}

We will now show how one can recover the spectral family $\ea$ of $A
\in \hr$ from the observable function $f_{A}$. This gives us the
decisive hint for the proof that to each abstract observable function 
$f : \dr \to \RR$ there is a unique $A \in \hr$ with $f = f_{A}$.
    
\begin{lemma}\label{ob16}
Let $ f_{A} : \dr \to  \RR$ be an observable function and let $\ea$ 
be the spectral family corresponding to $A$. If $\gl \in im f$, then 
\[
     \kj_{\gl} = \{ P \in \pr \mid \exists \  \mu > \gl : P ≥
     \ea_{\mu} \}.  
\]
$\kj_{\gl} = H_{\eal}$ if and only if $\ea$ is constant on some
interval $[\gl, \gl + \delta ]$. Moreover
\[
    \eal = \inf \kj_{\gl}.
\]
\end{lemma}

\begin{theorem}\label{ob17}
    Let $f : \dr \to \RR$ be an abstract observable function. Then
    there is a unique $A \in \hr$ such that $f = f_{A}$.
\end{theorem}
The {\bf proof} proceeds in three steps. In the first step we
construct from the abstract observable function $f$ an increasing family
$(\el)_{\gl \in im f}$ in $\pr$ and show in a second step that this
family can be extended to a spectral family in $\rr$. Finally, in the 
third step, we show that the selfadjoint operator $A \in
\rr$ corresponding to that spectral family has observable function
$f_{A} = f$ and that $A$ is uniquely determined by $f$.
We present here only the first two steps and omit the third because it
is rather technical.

\paragraph{Step 1}
    
Let $\gl \in im f$ and let $\jl \in \dr$ be the smallest dual ideal
such that $f(\jl) = \gl$. In view of lemma \ref{ob16} we have no
choice than to define
\[
    \el := \inf \jl.
\]
\begin{lemma}\label{ob18}
    The family $(\el)_{\gl \in im f}$ is increasing.
\end{lemma}
\emph{Proof:} Let $\gl, \mu \in im f, \ \gl < \mu$. Then
\begin{eqnarray*}
    f(\jm) & = & \mu  \\
     & = & max(\gl, \mu)   \\
     & = & max(f(\jl), f(\jm))  \\
     & = & f(\jl \cap \jm).
 \end{eqnarray*}
Hence, by the minimality of $\jm$,
\[
    \jm \tm \jl \cap \jm \tm \jl
\]
and therefore $\el ≤ \emm$. \ \ $\Box$

\begin{lemma}\label{ob19}
    $f$ is monotonely continuous, i.e. if $(\kj_{j})_{j \in J}$ is an 
    increasing net in $\dr$ then 
    \[
	f(\bigcup_{j \in J}\kj_{j}) = \lim_{j}f(\kj_{j}).
    \]
\end{lemma}
\emph{Proof:} Obviously $\kj := \bigcup_{j \in J}\kj_{j} \in \dr$. 
As $f$ is decreasing, $f(\kj) ≤ f(\kj_{j})$ for all $j \in J$ and
$(f(\kj_{j})_{j \in J}$ is a decreasing net of real numbers. Hence
\[
    f(\kj) ≤ \lim_{j}f(\kj_{j}).
\]
Let $\eps > 0$. Because $f$ is upper semicontinuous there is $P \in
\kj$ such that $f(\kI) < f(\kj) + \eps$ for all $\kI \in \dpr$. Now $P
\in \kj_{k}$ for some $k \in J$ and therefore
\[
     \lim_{j}f(\kj_{j}) ≤ f(\kj_{k}) < f(\kj) + \eps,
\]
which shows that also $\lim_{j}f(\kj_{j}) ≤ f(\kj)$ holds. \ \
$\Box$

\begin{corollary}\label{ob20}
    The image of an abstract observable function is compact.
\end{corollary}
\emph{Proof:} Because $\{I\} \tm \kj$ for all $\kj \in \dr$ we have $f
≤ f(\{I\})$ on $\dr$.\\
If $\gl, \mu \in im f$ and $\gl < \mu$ then $\jm \tm \jl$, hence
$\bigcup_{\gl \in im f}\jl$ is a dual ideal and therefore contained in
a maximal dual ideal $\frb \in \dr$. This shows $f(\frb) ≤ f$ on $\dr$
and consequently $im f$ is bounded. Let $\gl \in \overline{im f}$.
Then there is an increasing sequence $(\mu_{n})_{n \in \NN}$ in $im f$
converging to $\gl$ or there is a decreasing sequence $(\mu_{n})_{n \in
\NN}$ in $im f$ converging to $\gl$. In the first case we have
$\kj_{\mu_{n + 1}} \tm \kj_{\mu_{n}}$ for all $n \in \NN$ and
therefore for $\kj := \bigcap_{n}\kj_{\mu_{n}} \in \dr$
\[
    f(\kj) = \sup_{n}f(\kj_{\mu_{n}}) = \sup_{n}\mu_{n} = \gl.
\] 
In the second case we have $\kj_{\mu_{n}} \tm \kj_{\mu_{n + 1}}$
for all $n \in \NN$ and therefore $\kj := \bigcup_{n}\kj_{\mu_{n}} \in \dr$.
Hence
\[
    f(\kj) = \lim_{n}f(\kj_{\mu_{n}}) = \lim_{n}\mu_{n} = \gl.
\]
Therefore $\gl \in im f$ in both cases, i.e. $im f$ is also closed.
\ \ $\Box$

\paragraph{Step 2}

We will now extend $(\el)_{\gl \in im f}$ to a spectral family
$\ef := (\el)_{\gl \in \RR}$. In defining $\ef$ we have of course in
mind that the spectrum of the selfadjoint operator $A$ corresponding
to $\ef$ should coincide with $im f$. This forces us to define $\el$
for $\gl \notin im f$ in the following way. For $\gl \notin im f$ let
\[
    S_{\gl} := \{ \mu \in im f | \ \mu < \gl \}.
\]  
Then we define
\[
    \el :=
    \begin{cases}
	0     &  \text{if} \ \  S_{\gl} = \emptyset  \\
	E_{sup \  S_{\gl}}    &   \text{otherwise}.
    \end{cases}    
\]
Note that $f(\{ I \}) = max \ im f$ and that $\kj_{f(\{ I \})} = \{ I
\}$.
\begin{lemma}\label{ob21}
    $\ef$ is a spectral family.
\end{lemma}
\emph{Proof:} The only remaining point to prove is that $\ef$ is
continuous from the right, i.e. that $\el = \We_{\mu > \gl}\emm$ for
all $\gl \in \RR$. This is obvious if $\gl \notin im f$ or if there is
some $\gd > 0$ such that $]\gl, \gl + \gd[ \cap im f =\emptyset$.
Therefore we are left with the case that there is a strictly decreasing
sequence $(\mu_{n})_{n \NN}$ in $im f$ converging to $\gl$. For all $n
\in \NN$ we have $f(\kj_{\mu_{n}}) > f(\kj_{\gl})$ and therefore
$\kj_{\mu_{n}} \tm \kj_{\gl}$. Hence $\bigcup_{n}\kj_{\mu_{n}} \tm
\kj_{\gl}$ and 
\[
     f(\bigcup_{n}\kj_{\mu_{n}}) = \lim_{n}f(\kj_{\mu_{n}}) = \gl
\]
implies $\bigcup_{n}\kj_{\mu_{n}} = \kj_{\gl}$ by the minimality of
$\kj_{\gl}$. If $P \in \kj_{\gl}$ then $P \in \kj_{\mu_{n}}$ for some 
$n$ and therefore $E_{\mu_{n}} ≤ P$. This shows $\We_{\mu > \gl}\emm ≤
P$. As $P \in \kj_{\gl}$ is arbitrary we can conclude that $\We_{\mu >
\gl}\emm ≤ \el$. The reverse inequality is obvious. \ \ $\Box$ \\

The theorem confirms that there is no difference between ``abstract'' 
and ``concrete'' observable functions and therefore we will speak
generally of observable functions. \\

Let $\por$ denote the set of nonzero projections in $\rr$.
We will now show that observable functions can be characterized as
functions $ \por \to \RR$ that satisfy a ``continuous join condition''.
Note that for an arbitrary family $(P_{k})_{k \in \KK}$ in $\por$ we
have
\[
    \bigcap_{k \in \KK}H_{P_{k}} = H_{\Ve_{k \in \KK}P_{k}}. 
\]
If $f : \dr \to \RR$ is an observable function then the intersection
property implies
\[
     f(H_{\Ve_{k \in \KK}P_{k}}) = \sup_{k \in \KK}f(H_{P_{k}}).
\]
This leads to the following 

\begin{definition}\label{ob22}
    A bounded function $r : \por \to \RR$ is called completely increasing if
    \[
	 r(\Ve_{k \in \KK}P_{k}) = \sup_{k \in \KK}r(P_{k})
    \]
    for every family $(P_{k})_{k \in \KK}$ in $\por$.
\end{definition}

Note that it is sufficient to assume in the foregoing definition that 
$r$ is bounded from below because $r(I)$ is an upper bound, in fact
the maximum, for an arbitrary increasing function $r : \por \to \RR$. 
\\
Because of the natural bijection $P \tto H_{P}$ between $\por$ and the 
set $\dprr$ of principle dual ideals of $\pr$ each observable function
$f : \dr \to \RR$ induces by restriction a completely increasing function
$r_{f}$: 
\[
     \all \ P \in \por : \ r_{f}(P) := f(H_{P}).
\]
Conversely, each completely increasing function
on $\por$ induces an observable function so that we get a one to one
correspondence between observable functions and completely increasing 
functions. \\

\begin{definition}\label{ob23}
    Let $r : \por \to \RR$ be a completely increasing function. Then we
    define a function $f_{r} : \dr \to \RR$ by
    \[
	 \all \ \kj \in \dr : \ f_{r}(\kj) := \inf_{P \in \kj}r(P).
    \]
\end{definition}
It is obvious that
\[
    \all \ P \in \por : \ f_{r}(H_{P}) = r(P)
\]
holds.

\begin{proposition}\label{ob24}
    The function $f_{r} : \dr \to \RR$ induced by the completely increasing
    function $r : \por \to \RR$ is an observable function.
\end{proposition}

We have formulated theorem \ref{ob17} and the
characterization of observable functions by completely increasing 
functions in the category of von Neumann algebras. A simple
inspection of the proofs shows that we have used the fact that the
projection lattice $\pr$ of a von Neumann algebra $\rr$ is a complete
orthomodular lattice. Therefore we can translate theorem
\ref{ob17} to the category of complete orthomodular lattices
in the following way:
\begin{theorem}\label{ob25}
    Let $\LL$ be a complete orthomodular lattice and let $f : \dl \to \RR$
    be an abstract observable function. Then there is a unique
    spectral family $E$ in $\LL$ such that $f = f_{E}$.
\end{theorem}

Finally, we like to present a characterization of those bounded
continuous functions $f : \qr \to \RR$ that are observable. \\

Let $f : \qr \to \RR$ be a bounded continuous function. Because of 
\[
    H_{P} = \bigcap \qpr
\]
for all $P \in \por$ it is natural to define
\[
    r(P) := \sup \{ f(\frb) | \frb \in \qpr \}.
\]
Suppose that $r : \por \to \RR$ is completely increasing. (If $\rr$ is
abelian, then this assumption is automatically satisfied.) Let $f_{r}
: \dr \to \RR$ be the corresponding observable function.

\begin{lemma}\label{ob26}
    $f$ coincides with the restriction of $f_{r}$ to $\qr$.
\end{lemma}
\emph{Proof:} We have to show that
\[
    \all \ \frb \in \qr : \ f(\frb) = \inf_{P \in \frb}r(P) 
\]
holds. \\
From the definition of $r$ we see that $f(\frb) ≤ m := \inf_{P \in
\frb}r(P)$. Let $\eps > 0$. Because $f$ is continuous there is $P_{0} 
\in \frb$ such that $f(\frc) < f(\frb) + \eps$ on $\kQ_{P_{0}}(\rr)$. 
Hence 
\[
    m ≤ r(P_{0}) = \sup f(\kQ_{P_{0}}(\rr)) ≤ f(\frb) + \eps.
\]   
This shows $m ≤ f(\frb)$. \ \ $\Box$

Finally, we can formulate a criterion for the observability of a
continuous function $f : \qr \to \RR$:

\begin{theorem}\label{ob27}
    Let $\rr$ be a von Neumann algebra. Then a bounded continuous
    function $f : \qr \to \RR$ is an observable function if and only
    if the induced function
    \[
        \begin{array}{cccc}
            r_{f} : & \por & \to & \RR  \\
             & P & \tto & \sup_{\frb \in \qpr}f(\frb)
        \end{array}
    \]
    is completely increasing.
\end{theorem}

\subsection{Classical Observables}

In the previous subsection we have seen that selfadjoint elements 
$A$ of a von Neumann algebra $\rr$ correspond to certain continuous
real valued functions $f_{A}$ on the Stone spectrum $\qr$ of $\pr$.\\
  
Now we will show that continuous real valued 
functions on a Hausdorff space $M$ can be described by 
spectral families with values in the complete lattice $\kT(M)$ of 
open subsets of $M$. These spectral families $\gs : \RR \to \kT(M)$
can be characterized abstractly by a certain property of 
the mapping $\gs$. Thus also a classical observable has a ``quantum
mechanical'' description. This shows that classical and quantum
mechanical observables are on the same structural footing: either as
functions or as spectral families. Similar results hold for functions 
on a set $M$ that are measurable with respect to a $\gs$-algebra of 
subsets of $M$. Due to our time limit, we confine ourselves to the
statement of the main result. \\

We would like to begin with some simple examples:
  \begin{example}\label{ob28}
  The following settings define spectral families $\gs_{id}, 
  \gs_{abs}, \gs_{ln}, \gs_{step}$ in $\kT(\RR)$:
  \begin{eqnarray}
		    \gs_{id}(\gl) & := & ]-\infty, \gl[,
		    \label{1}  \\
		    \gs_{abs}(\gl) & := & ]-\gl, \gl[
		    \label{2}  \\
		    \gs_{ln}(\gl) & := & ]-\exp(\gl), \exp(\gl)[
		    \label{3}  \\
		    \gs_{step}(\gl) & := & ]-\infty, \lfloor\gl\rfloor[
		    \label{4}
  \end{eqnarray}
where $\lfloor\gl\rfloor$ denotes the ``floor of $\gl \in \RR$'':
  \begin{displaymath}
		    \lfloor\gl\rfloor = \max \{n \in \ZZ \mid n \leq \gl \}.
  \end{displaymath}
  \end{example}
  The names of these spectral families sound somewhat crazy at the 
  moment, but we will justify them soon.
  
  In close analogy to the case of spectral families in the 
  lattice $\LL(\kh)$, each spectral family in $\kT(M)$ induces a 
  function on a subset of $M$.

  \begin{definition}\label{ob29}
  Let $\gs : \RR \to \kT(M)$ be a spectral family in $\kT(M)$. Then
  \begin{displaymath}
	    \kD(\gs) := \{x \in M \mid \exists \ \gl \in \RR : \ x \notin 
		    \gs(\gl) \}
  \end{displaymath}
  is called the {\bf admissible domain of $\gs$}.
  \end{definition}
  
  Note that 
  \[
      \kD(\gs) = M \smm \bigcap_{\gl \in \RR}\gs(\gl).
  \]
  
  \begin{remark}\label{ob30}
  The admissible domain $\kD(\gs)$ of a spectral family $\gs : \RR 
  \to \kT(M)$ is dense in $M$.
  \end{remark}
  
  On the other hand it may happen that $\kD(\gs) \ne M$. The 
  spectral family $\gs_{ln}$ is a simple example:
  \begin{displaymath}
	    \forall \ \gl \in \RR : \ 0 \in \gs_{ln}(\gl).
  \end{displaymath}
  
  Each spectral family $\gs : \RR \to \kT(M)$ induces a function 
  $f_{\gs} : \kD(\gs) \to \RR$:
  
  \begin{definition}\label{ob31}
  Let $\gs : \RR \to \kT(M)$ be a spectral family with admissible 
  domain $\kD(\gs)$. Then the function $f_{\gs} : \kD(\gs) \to 
  \RR$, defined by
  \begin{displaymath}
      \forall \ x \in \kD(\gs) : \ f_{\gs}(x) := \inf \{\gl \in \RR 
		    \mid x \in \gs(\gl) \},
  \end{displaymath}
  is called the {\bf function induced by $\gs$}.
  \end{definition}

In complete analogy to the operator case we define the spectrum of 
a spectral family $\gs$:

\begin{definition}\label{ob32}
Let $\gs : \RR \to \kT(M)$ be a spectral family. Then
\begin{displaymath}
	  R(\gs) := \{\gl \in \RR \mid \gs \  \text{is constant on a 
		  neighborhood of}\  \gl \}
\end{displaymath} 
is called the {\bf resolvent set} of $\gs$, and
\begin{displaymath}
		  sp(\gs) := \RR \setminus R(\gs)
\end{displaymath}
is called the {\bf spectrum} of $\gs$.
\end{definition}

Obviously $sp(\gs)$ is a closed subset of $\RR$.

\begin{proposition}\label{ob33}
Let $f_{\gs} : \kD(\gs) \to \RR$ be the function induced by the 
spectral family $\gs : \RR \to \kT(M)$. Then
\begin{displaymath}
		  sp(\gs) = \overline{im f_{\gs}}.
\end{displaymath}
\end{proposition}

The functions induced by our foregoing examples are
\begin{eqnarray}
		  f_{\gs_{id}}(x) & = & x
		  \label{1}  \\
		  f_{\gs_{abs}}(x) & = & |x|
		  \label{2}  \\
		  f_{\gs_{ln}}(x) & = & \ln{|x|} \quad  \text{and} \ \kD(\gs_{ln}) = \RR 
		  \setminus\{0\}
		  \label{3}  \\
		  f_{\gs_{step}} & = & \sum_{n \in \ZZ}n\chi_{[n, n + 1[}
		  \label{4}
\end{eqnarray}
There is a fundamental difference between the spectral families 
$\gs_{id}, \gs_{abs}, \gs_{ln}$ on the one side and $\gs_{step}$ 
on the other. The function induced by $\gs_{step}$ is not 
continuous. This fact is mirrored in the spectral families: the 
first three spectral families have the property
\begin{displaymath}
	  \forall \ \gl < \mu : \ \overline{\gs(\gl)} \tm \gs(\mu).
\end{displaymath}
Obviously $\gs_{step}$ fails to have this property.

\begin{definition}\label{ob34}
A spectral family $\gs : \RR \to \kT(M)$ is called {\bf 
continuous} if
\begin{displaymath}
	  \forall \ \gl < \mu : \ \overline{\gs(\gl)} \tm \gs(\mu)
\end{displaymath}
holds.
\end{definition}      

\begin{remark}\label{ob35}
The admissible domain $\kD(\gs)$ of a continuous spectral 
family $\gs : \RR \to \kT(M)$ is an open (and dense) subset of $M$.
\end{remark}

\begin{remark}\label{ob36}
If $\gs : \RR \to \kT(M)$ is a continuous spectral family, then for 
all $\gl \in \RR \qquad \gs(\gl)$ is a regular open set, i.e.
$\gs(\gl)$ is the interior of its closure.
\end{remark}

The importance of continuous spectral families becomes manifest 
in the following

\begin{theorem}\label{ob37}
Let $M$ be a Hausdorff space. Then every continuous function $f : 
M \to \RR$ induces a continuous spectral family $\gs_{f} : \RR 
\to \kT(M)$ by
\begin{displaymath}
  \forall \ \gl \in \RR : \ \gs_{f}(\gl) := int(\overset{-1}{f}(]-\infty,
  \gl])).
\end{displaymath}
The admissible domain $\kD(\gs_{f})$ equals $M$ and the function 
$f_{\gs_{f}} : M \to \RR$ induced by $\gs_{f}$ is $f$. 
Conversely, if $\gs : \RR \to \kT(M)$ is a continuous spectral 
family, then the function
\begin{displaymath}
		  f_{\gs} : \kD(\gs) \to \RR
\end{displaymath}
induced by $\gs$ is continuous and the induced spectral family 
$\gs_{f_{\gs}}$ in $\kT(\kD(\gs))$ is the restriction of $\gs$ to 
the admissible domain $\kD(\gs)$:
\begin{displaymath}
  \forall \gl \in \RR : \ \gs_{f_{\gs}}(\gl) = \gs(\gl) \cap \kD(\gs). 
\end{displaymath}
\end{theorem}

One may wonder why we have defined the function, that is induced by a
spectral family $\gs$, on $M$ and not on the Stone spectrum $\qm$. A
quasipoint $\frb \in \qm$ is called \emph{finite} if $\bigcap_{U \in
\frb}\overline{U} \ne \emptyset$. If $\frb$ is finite, then this
intersection consists of a single element $x_{\frb} \in M$, and we call
$\frb$ a quasipoint \emph{over $x_{\frb}$}. Note that for a compact space
$M$, all quasipoints are finite. Moreover, one can show that for
compact $M$, the mapping $pt : \frb \tto x_{\frb}$ from $\qm$ onto $M$
is continuous and \emph{identifying}.

\begin{remark}\label{ob38}
    Let $\gs : \RR \to \kT(M)$ be a continuous spectral family and let
    $x \in \kD(\gs)$. Then for all quasipoints $\frb_{x} \in \qm$ over
    $x$ we have
    \[
	f_{\gs}(\frb_{x}) = f_{\gs}(x).
    \]
\end{remark}
Therefore, if $M$ is compact, it makes no difference whether we define
$f_{\gs}$ in $M$ or in $\qm$.  
\pagebreak

\section{Presheaves Again: Quantum Observables as Global Sections}
\label{pp}

The abstract characterization of (quantum) observable functions leads 
to a natural definition of restricting selfadjoint elements of a von
Neumann algebra $\rr$ to a subalgebra $\mm$. Again we denote a
completely increasing function on $\por$ and the corresponding observable
function (on $\qr$ or $\dr$) by the same letter and speak simply of an
observable function. We denote the set of observable functions for
$\rr$ by $\orr$. Obviously we have

\begin{remark}\label{pp1}
    Let $\mm$ be a von Neumann subalgebra of a von Neumann algebra
    $\rr$ and let $f : \por \to \RR$ be an observable function. Then
    the restriction
    \[
	\gr_{\mm}f := f_{\mid_{\pom}}
    \]   
is an observable function for $\mm$. It is called the restriction of
$f$ to $\mm$.    
\end{remark}

This definition is absolutely natural. However, if $A$ is a
selfadjoint operator in $\rr$ then the observable function $f_{A} :
\por \to \RR$ corresponding to $A$ is a rather abstract encoding
of $A$. So before we proceed we will describe the restriction map
\[
    \begin{array}{ccc}
	\gr_{\mm} : \orr & \to & \omm  \\
	f_{A} & \tto & \gr_{\mm}f_{A}
    \end{array}
\] 
in terms of spectral families. \\

To this end we define

\begin{definition}\label{pp2}
    Let $\kf$ be a filterbase in $\por$. Then
    \[
	C_{\rr}(\kf) := \{ Q \in \por \ | \ \ex \ P \in \kf : \ P ≤ Q 
	\}
    \]
    is called the cone over $\kf$ in $\rr$.
\end{definition}
Clearly $C_{\rr}(\kf)$ is a dual ideal and it is easy to see that it
is the \emph{smallest dual ideal that contains $\kf$.} A dual ideal
$\kI \in \dm$ is, in particular, a filterbase in $\pr$, so
$C_{\rr}(\kI)$ is well defined. 

\begin{proposition}\label{pp3}
    Let $f \in \orr$. Then 
    \[
	(\gr_{\mm}f)(\kI) = f(C_{\rr}(\kI))
    \]
    for all $\kI \in \dm$.
\end{proposition}

\begin{definition}\label{pp4}
    For a projection $Q$ in $\rr$ let
    \[
	c_{\mm}(Q) := \Ve \{ P \in \pmm \ | \ P ≤ Q \} \quad
	\text{and} \quad s_{\mm}(Q) := \We \{ P \in \pmm \ | \ P ≥ Q \}.
    \]
    $c_{\mm}(Q)$ is called the {\bf $\mm$-core}, $s_{\mm}(Q)$ the
    {\bf $\mm$-support} of $Q$.
\end{definition}

The $\mm$-support is a natural generalization of the notion of central
support which is the $\mm$-support if $\mm$ is the center of $\rr$. Note
that if $Q \notin \mm$ then $c_{\mm}(Q) < Q < s_{\mm}(Q)$.
The $\mm$-core and the $\mm$-support are related in a simple manner:

\begin{remark}\label{pp4a}
    $c_{\mm}(Q) + s_{\mm}(I - Q) = I$ for all $Q \in \pr$.
\end{remark}

\begin{lemma}\label{pp5}
    Let $E = (\el)_{\gl \in \RR}$ be a spectral family in $\rr$ and
    for $\gl \in \RR$ define
    \[
	(c_{\mm}E)_{\gl} := c_{\mm}(\el), \quad (s_{\mm}E)_{\gl} :=
	\We_{\mu > \gl}s_{\mm}(\emm).
    \]
    Then $c_{\mm}E := ((c_{\mm}E)_{\gl})_{\gl \in \RR}$ and
    $s_{\mm}E := ((s_{\mm}E)_{\gl})_{\gl \in \RR}$ are spectral
    families in $\mm$.
\end{lemma}

\begin{proposition}\label{pp6}
    Let $f \in \orr$ and let $E$ be the spectral family corresponding 
    to $f$. Then $c_{\mm}E$ is the spectral family corresponding to
    $\gr_{\mm}f$.
\end{proposition}
\emph{Proof:} Let $\kI$ be a dual ideal in $\pmm$. Then
$(\gr_{\mm}f)(\kI) = f(C_{\rr}(\kI))$ and
\begin{eqnarray*}
    f(C_{\rr}(\kI)) & = & \inf \{ \gl \ | \ \el \in C_{\rr}(\kI) \}  \\
     & = & \inf \{ \gl \ | \ \ex \ P \in \kI : \ P ≤ \el \}  \\
     & = & \inf \{ \gl \ | \ c_{\mm}(\el) \in \kI \}.
\end{eqnarray*}
Thus the assertion follows from theorem \ref{ob17}. \ \ $\Box$ \\

By theorem \ref{ob17}, the restriction map $\gr_{\mm} : \orr \to
\omm$ induces a restriction map
\[
\begin{array}{cccc}
    \gr_{\mm} : & \hr & \to & \mm_{sa}  \\
     & A & \tto & \gr_{\mm}A
\end{array}
\]
for selfadjoint operators. In particular, we obtain

\begin{corollary}\label{pp7}
    $\gr_{\mm}Q = s_{\mm}(Q)$ for all projections $Q$ in $\rr$.
\end{corollary}

The corollary shows that the restriction map $\gr_{\mm} : \hr \to
\mm_{sa}$ has the important property that it maps projections to
projections and acts as the identity on $\pmm$. It also shows that
in general $\gr_{\mm}$ is not linear: if $P, Q \in \pr$ such that $PQ 
=0$ then it is possible that $s_{\mm}(P)s_{\mm}(Q) \ne 0$ and
therefore $s_{\mm}(P + Q) \ne s_{\mm}(P) + s_{\mm}(Q)$.\\

Proposition \ref{pp6} and lemma \ref{pp5} suggest still 
another natural possibility for defining a restriction map $\gs_{\mm} 
: \hr \to \mm_{sa}$: if $\ea$ is the spectral family corresponding to
$A \in \hr$ then $\gs_{\mm}A$ is the selfadjoint operator defined by
the spectral family $s_{\mm}\ea$. \\

One can define a new partial order on $\hr$ by 
\[
    A ≤_{s} B \quad :\llra \quad f_{A} ≤ f_{B}. 
\]
On the level of spectral families, this can be expressed by
\[
    A ≤_{s} B \quad :\llra \quad \all \ \gl \irr : \ \ebl ≤ \eal.
\]
This is the reason for calling $≤_{s}$ the \emph{spectral order} on
$\hr$. It is coarser than the usual one, but is has the advantage  to 
turn $\hr$ to a \emph{boundedly complete lattice} (\cite{deg, ol}).\\ 

\begin{proposition}\label{pp8}
    Let $\mm$ be a von Neumann subalgebra of the von Neumann algebra
    $\rr$. Then, for all $A \in \hr$, we have
    \[
	\gs_{\mm}A = \Ve \{ B \in \mm_{sa} \ | \ B ≤_{s} A \}
    \]
    and 
    \[
	\gr_{\mm}A = \We \{ C \in \mm_{sa} \ | \ A ≤_{s} C \}, 
    \]
    where $\gs_{\mm}A, \gr_{\mm}A$ are considered as elements of 
    $\rr$ and $\Ve, \We$ denote the greatest lower bound and the
    least upper bound with respect to the spectral order.
\end{proposition}
This proposition shows that the two restriction mappings $\gr_{\mm}$
and $\gs_{\mm}$ from $\hr$ onto $\mm_{sa}$ are on an equal footing. It
is not difficult to determine the observable function $\gs_{\mm}f$
corresponding to $\gs_{\mm}A$, given the observable function $f$
corresponding to $A \in \hr$. \\
The restrictions $\gr_{\mm}A$ and $\gs_{\mm}A$ can be seen as coarse
grainings of $A$. We will demonstrate this in the abelian case: let
$\kaa$ be a von Neumann subalgebra of the abelian von Neumann algebra 
$\kbb$. We will show how the restriction maps $\kba : \kbb_{sa} \to
\kaa_{sa}$ and $\gs^\kbb_{\kaa} : \kbb_{sa} \to \kaa_{sa}$
act on observable functions $f : \qb \to \RR$ or, in other words, how the
Gelfand transformation behaves with respect to the restrictions $\kba$
and $\gs^\kbb_{\kaa}$.   

\begin{proposition}\label{pp9}
   Let $\kaa, \kbb$ be as above and let $f : \qb \to \RR$ be an observable
   function. Then we have for all $\gga \in \qa$: 
   \begin{enumerate}
       \item  [(i)] $(\kba f)(\gga) = \sup \{ f(\gb) \ | \ \gga \tm \gb
       \}$ and            
   
       \item  [(ii)] $(\gs^\kbb_{\kaa}f)(\gga) = \inf \{ f(\gb) \ | \
       \gga \tm \gb \}$.      
   \end{enumerate}
\end{proposition}

Now consider three von Neumann subalgebras $\kaa, \kbb, \kcc$ of $\rr$
such that $\kaa \tm \kbb \tm \kcc$. Then the corresponding restriction
maps $\kcb : \kcc_{sa} \to \kbb_{sa}, \ \kba : \kbb_{sa} \to \kaa_{sa}$
and $\kca : \kcc_{sa} \to \kaa_{sa}$ obviously satisfy
\begin{equation}
    \kca = \kba \circ \kcb \quad \text{and} \quad \gr^{\kaa}_{\kaa} =
    id_{\kaa_{sa}}.
    \label{eq:ps}
\end{equation}
The set $\frsr$ of all von Neumann subalgebras of $\rr$ is a lattice
with respect to the partial order given by inclusion. The meet of
$\kaa , \kbb \in \frsr$ is defined as the intersection,
\[
    \kaa \we \kbb := \kaa \cap \kbb,
\]
and the join as the subalgebra generated by $\kaa$ and $\kbb$:
\[
    \kaa \vee \kbb := (\kaa \cup \kbb)''.
\]
The join is a rather intricate operation. This can already be seen in
the most simple (non-trivial) example $lin_{\CC}\{I, P\} \vee
lin_{\CC}\{I, Q\}$ for two non-commuting projections $P, Q \in \rr$.
Fortunately we don't need it really. \\

The subset $\frAr \tm \frsr$ of all \emph{abelian} von Neumann
subalgebras of $\rr$ is also partially ordered by inclusion but it is 
only a \emph{semilattice}: the meet of two (in fact of an arbitrary
family of) elements of $\frAr$ always exists but the join does not in 
general. Both $\frsr$ and $\frAr$ have a smallest element, namely
$\mathbb{O} := \CC I$. However, unless $\rr$ is itself abelian, there
is no greatest element in $\frAr$. Anyway, $\frsr$ and $\frAr$ can be 
considered as the sets of objects of (small) \emph{categories} whose
morphisms are the inclusion maps. \\
In quantum physics the (maximal) abelian von Neumann subalgebras of
$\lh$ are called \emph{contexts}. We generalize this notion in the
following

\begin{definition}\label{pp10}
    The small category $\cor$, whose objects are the abelian von Neumann
    subalgebras of $\rr$ and whose morphisms are the inclusion maps,
    is called the {\bf context category of the von Neumann algebra
    $\rr$}.
\end{definition}

We define a \emph{presheaf} $\kO_{\rr}$ on the context category $\cor$
of $\rr$ by sending objects $\kaa \in \frAr$ to $\kO_{\rr}(\kaa) :=
\kaa_{sa}$ (or equivalently to $\oaa$) and morphisms $\kaa
\hookrightarrow \kbb$ to restrictions $\kba : \kbb \to \kaa$. This
gives a contravariant functor, i.e. a presheaf on $\cor$.
\begin{definition}\label{pp11}
    The presheaf $\kO_{\rr}$ is called the {\bf observable presheaf} of
    the von Neumann algebra $\rr$.
\end{definition}

Every observable function $f \in \orr$ induces a
family $(f_{\kaa})_{\kaa \in \frAr}$ of observable functions $f_{\kaa}
\in \oaa$, defined by $f_{\kaa} := \gr^{\rr}_{\kaa}f$. This family has
the following compatibility property:
\begin{equation}
    \all \ \kaa, \kbb \in \frAr : \ \gr^{\kaa}_{\kaa \cap \kbb}f_{\kaa}
	= \gr^{\kbb}_{\kaa \cap \kbb}f_{\kbb}.
    \label{eq:gs1}
\end{equation}
$(f_{\kaa})_{\kaa \in \frAr}$ is therefore a \emph{global section} of 
the presheaf $\kO_{\rr}$ in the following general sense.

\begin{definition}\label{pp12}
    Let ${\bf C}$ be a category and $\kS : {\bf C} \to {\bf Set}$ a presheaf,
    i.e. a contravariant functor from ${\bf C}$ to the category ${\bf
    Set}$ of sets. A global section of $\kS$ assigns to every object 
    $a$ of ${\bf C}$ an element $\gs(a)$ of the set $\kS(a)$ such that
    for every morphism $\gf : b \to a$ of ${\bf C}$
    \[
	\gs(b) = \kS(\gf)(\gs(a))
    \]
    holds.
\end{definition}

In the case of the observable presheaf $\kO_{\rr}$ there are plenty of
global sections because each $A \in \hr$ induces one. Here the natural
question arises whether all global sections of $\kO_{\rr}$ are induced
by selfadjoint elements of $\rr$. This is certainly not true if the
Hilbert space $\kh$ has dimension two. For in this case the
constraints \ref{eq:gs1} are void and therefore \emph{any} function on
the complex projective line defines a global section of $\kO_{\lh}$.
But Gleason's (or Kochen-Specker's) theorem teaches us that the
dimension two is something peculiar. One can show, however, that the
phenomenon, that there are global sections of $\kO_{\rr}$ that are not 
induced by selfadjoint elements of $\rr$, is not restricted to the
dimension two. \\

Therefore, if one takes contextuality in quantum physics serious, it
is natural to generalize the notion of quantum observable:

\begin{definition}\label{pp13}
    Let $\rr$ be a von Neumann algebra. The global sections of the observable
    presheaf $\kO_{\rr}$ are called {\bf contextual observables}.
\end{definition}

Contextual observables can be characterized as certain functions on
$\por$:
\pagebreak
\begin{proposition}\label{gs7}
    Let $\rr$ be a von Neumann algebra. There is a one-to-one
    correspondence between global sections of the observable presheaf
    $\kO_{\rr}$ and functions $f : \por \to \RR$ that satisfy
    \begin{enumerate}
	\item  [(i)] $f(\Ve_{k \ikk}P_{k}) = \sup_{k \ikk}f(P_{k})$ for all
	{\bf commuting} families $(P_{k})_{k \ikk}$ in $\por$,
    
	\item  [(ii)] $f_{|_{\por \cap \kaa}}$ is bounded for all
	$\kaa \in \frAr$.
    \end{enumerate}
\end{proposition}
\emph{Proof:} Let $(f_{\kaa})_{\kaa \in \frAr}$ be a global section of 
$\kO_{\rr}$. Then the functions $f_{\kaa} : \por \cap \kaa \ (\kaa
\in \frAr)$ can be glued to a function $f : \por \to \RR$:\\
Let $P \in \por$ and let $\kaa$ be an abelian von Neumann subalgebra
of $\rr$ that contains $P$. Then 
\[
    f(P) := f_{\kaa}(P)
\]
does not depend on the choice of $\kaa$. Indeed, if $P \in \kaa \cap
\kbb$, then $f_{\kaa}(P) = f_{\kbb}(P)$ by the compatibility property 
of global sections. It is obvious that $f$ satisfies properties $(i)$ 
and $(ii)$. \\
If, conversely, a function $f : \por \to \RR$ with the properties
$(i)$ and $(ii)$ is given and if $\kaa$ is an abelian von Neumann subalgebra
of $\rr$, then $f_{\kaa} := f_{|_{\por \cap \kaa}}$ is a completely
increasing function. The family $(f_{\kaa})_{\kaa \in \frAr}$ is then,
by construction, a global section of $\kO_{\rr}$. \ \ $\Box$ \\

\pagebreak

\end{document}